\documentclass{llncs}

\usepackage{color}

\usepackage{epsfig}
\usepackage{amsfonts}
\usepackage{latexsym}
\usepackage{amssymb}
\usepackage{amsmath}

\newcommand{\q}{\phantom{0}}
\newcommand{\qq}{\phantom{00}}
\newcommand{\dqq}{\phantom{.00}}

\begin{document}

\mainmatter

\title{Engineering Relative Compression of Genomes}

\author{
  Szymon Grabowski\inst{1}\thanks{Supported by Polish Ministry of Science 
and Higher Education grant N N516 441938}
  \and
  Sebastian Deorowicz\inst{2}\thanks{Supported by Polish Ministry of Science 
and Higher Education grant N N516 441938}
  \institute{
    Technical University of {\L}\'od\'z, Computer Engineering Department\\
    Al.\ Politechniki 11, 90--924 {\L}\'od\'z, Poland\\
    \email{sgrabow@kis.p.lodz.pl}
    \and
    Institute of Informatics, Silesian University of Technology\\
    Akademicka 16, 44-100 Gliwice, Poland\\
    \email{sebastian.deorowicz@polsl.pl}
  }
}

\maketitle

\begin{abstract}
Technology progress in DNA sequencing boosts the genomic database
growth at faster and faster rate.
Compression, accompanied with random access capabilities, is the key
to maintain those huge amounts of data.
In this paper we present an LZ77-style compression scheme for 
relative compression of multiple genomes of the same species.
While the solution bears similarity to known algorithms, it 
offers significantly higher compression ratios 
at compression speed over a order of magnitude greater.
One of the new successful ideas is augmenting the reference 
sequence with phrases from the other sequences, making more 
LZ-matches available.

\smallskip

\end{abstract}

\section{Introduction}

The old saying ``Time is money'' remains true in IT systems, where
time savings are often accomplished by means of data compression
techniques, especially if they are adopted to the characteristics of
processed data and coupled with procedures for fast search or
random access directly over the compressed representation.
Data compression not only saves the storage but also reduces
transmission times and can even let extracting particular pieces
of information faster.
The last phenomenon has long been observed in database communities,
and can be attributed to the fact that
more compact form reduces the number of memory accesses,
particularly important in the hierarchical computer architectures
(disks, RAM, caches of different levels, CPU registers),
so common these days.

One of the main tasks of bioinformatics is to collect and analyze
huge genomic data, typically obtained with sequencing methods.
Rapid development in DNA sequencing technologies led to drastic
growth of data publicly available in sequence databases, e.g.,
GenBank at NCBI or 1000 Genomes project, to name a few.
To illustrate this progress, we note that the first draft
sequence of the human genome was published in 2001, and now,
a decade later, the cost of acquiring a single (individual) human
genome is below \$10,000 and is still expected to drop significantly
in the next few years.
Storing genetic sequences of many individuals of the same species
promises new discoveries for the whole field of biology,
and low cost of acquiring an individual human genome gives way to
``personalized medicine'', making use of one's individual genetic profile.

It should be stressed that DNA sequences within the same species
are both large and highly repetitive.
For example, only about 0.1\% of the 3\,GB human genome is specific;
the rest is common to all humans.
This poses interesting challenges to efficient storage and fast access to
those data.
Most classic data compression techniques fail to recognize this
tremendous redundancy, simply because finding matches with e.g.
an LZ77 variant with a sliding window would require a multi-gigabyte
buffer, not counting the match-finding structures.
Using a context-based statistical coding (e.g., PPM) may efficiently exploit
the repetitions only if the considered context is long enough, otherwise
the context statistics are polluted with ``accidental'' matches.
This phenomenon is specific for genomic data, which are almost incompressible
within a single individual, which also means that short subsequences,
e.g., shorter than 10 symbols, are very likely to occur many times in
seemingly random places.
On the other hand, maintaining a high-order statistical model over a large
collection of data may be prohibitive due to enormous memory requirements.

Interestingly, also most of the DNA-specialized compressors from the
literature are not appropriate to handle modern genomic databases.
There are a number of reasons for that:
(\emph{i}) most of them focus on compression ratio, not on compression and
decompression speed or the memory use during the compression process;
from those reasons those cannot be practically run on sequences larger
than, say, several megabytes;
(\emph{ii}) considering the compression ratio, most effort has been put to
succinctly represent a single genome (which is believed to be almost
incompressible anyway, hence only tiny improvements were at the stake),
not to be particularly efficient in detecting inter-genome redundancy;
(\emph{iii}) most of the algorithms do not support extracting a range of symbols
from the middle of the compressed stream.
Only since around 2009 we can observe a surge of interest in practical,
multi-sequence oriented, DNA compressors, usually coupled with random access
capabilities and sometimes also offering indexed search.
We are going to take a close look at those solutions in the next section.

A different, even if related, problem is compression of large sets of
sequence reads (usually obtained from Solid, Illumina or 454 sequencers). 
Those data contain three streams: Title lines, DNA symbols and quality scores, 
and in the main stream it is possible to look for LZ77 matches only over 
those DNA symbols for which the quality scores are high enough 
(i.e., in non-noisy areas) to improve compression ratio, 
compression speed and reduce memory requirements~\cite{DG11}.

In this work, we propose an algorithm for effective compression of 
multiple genomic (DNA) sequences of the same species
and show experimental results suggesting its supremacy over existing solutions.
Although the general framework of relative LZ77-style compression is not new 
in this context, we add some new ideas to the existing algorithms.

\section{Related work}

The first DNA-specific compressor was BioCompress~\cite{GT93},
presented in 1993 by Grumbach and Tahi.
It was basically an LZ77 compression scheme, but also with support
for reverse complement repeats.
Since then, more than ten other compressors have been published,
and the new ideas involved encoding approximate repeats,
order-2 arithmetic encoding for literals,
forming probability distributions for each symbol on the basis of
approximate partial matches,
combining predictions from a panel of experts,
off-line substitution of repeated substrings, and more
(the references can be found in the survey~\cite{GSU09}).
Unfortunately, most of the algorithms presented in the literature
are computationally intensive and have been tested only
on small datasets (on the order of a few MB's or less).\footnote{In
case of many papers on DNA compression, much longer sequences,
to experiment with, were not yet available at their publication time.}

Probably the first DNA compressors running at acceptable speed
were the variants proposed by Manzini and Rastero~\cite{MR04}.
They resigned from searching for approximate repeats, as a major
culprit in causing the slowness of many other solutions in this area.
Instead, their algorithms are able to detect exact and
reverse complement repeats and are (implicitly) efficient on
approximate matches thanks to making use of frequent regularities
in match locations of successive matches.
Moreover, the stronger (and slower) variants, dna2 and dna3, apply
order-2 or order-3 arithmetic encoding of some components in their
sequence parsing.

In 2009, the first algorithms focused on compressing DNA sequences
from the same species have appeared~\cite{CLLX09,BWB09}, followed soon
by more mature proposals~\cite{MNSV10,CFMPN10,KurPZ10,KurPZ11,KN10,KN11}.
We present them in the following paragraphs.

M\"akinen et al.~\cite{MNSV10} added new functionalities to compressed
DNA sequences: \emph{display} (which can also be called the random
access functionality) returning the substring specified by its start 
and end position, \emph{count} telling the number of times the given pattern 
occurs in the text, and \emph{locate} listing the positions of the pattern 
in the text.
Although those operations are not new in full-text indexes
(possibly also compressed), the authors noticed that the existing
general solutions, paying no attention to long repeats in the input
(for the survey of compressed full-text indexes, see~\cite{NM07}),
are not very effective here and they proposed novel {\em self-indexes}
for the considered problem.

Claude et al.~\cite{CFMPN10} pointed out that the full-text indexes
from~\cite{MNSV10}, albeit fast in counting, are relatively slow
in extracting the locations of the matches, a feature shared by
all compressed indexes based on the {\em Burrows--Wheeler
transform}~\cite{BW94,NM07}.
They proposed two schemes.
One is an inverted index on $q$-grams, tailored to the repetitive
nature of the input data by using a strong LZ77-style compressor
(7-zip) on byte-encoded differential posting lists.
The text itself, i.e., the concatenation of all individual sequences,
is compressed with a popular grammar-based algorithm Re-Pair~\cite{LM00},
capable of extracting arbitrary substrings fast.
The other scheme is a purely grammar-based self-index.
The inverted index offers excellent space-time tradeoffs (on real data,
not in the worst case), but can basically work with substrings of
fixed length $q$ (we believe it is possible to heuristically adapt
this scheme to any substring length, but this issue is not discussed
in the cited work).
The grammar-based self-index is more elegant and can work with any
substring length, but uses significantly more space, is slower
and needs a large amount of RAM in the index build phase.

Kuruppu et al.~\cite{KurPZ10}, in their earlier work, propose a
surprisingly simple yet quite efficient compression scheme
with random access (it is not an index though).
They choose one of the sequences as the reference sequence and
compress it with a self-index (in the experiments in the cited work,
however, a general-purpose compressor, 7-zip, with no random access
capabilities, was used).
The other sequences are greedily parsed into LZ-factors of the
reference sequence.
Some extra data structure is added to provide random access.

The work~\cite{KurPZ11} from the same team is a follow-up paper, 
presenting a stronger LZ77-style algorithm. 
In this work, however, the authors' implementation is not
augmented with extra data enabling random access.
Still, it is mentioned how it can be (easily) added and we
are convinced the extra penalty in space will be small.
As their algorithm is a departure point of our proposal, 
we describe it more extensively in the next section.

Kreft and Navarro~\cite{KN10} presented an innovative LZ77 variant 
called LZ-End enabling extraction of arbitrary phrases in optimal 
time complexity.
While this algorithm has many applications, one of the immediate ones 
is for compressing DNA sequences, combining quite good compression 
ratios with very fast access.

In a recent work from the same authors~\cite{KN11}, a self-index 
based on the LZ-End idea is developed, with less than 3 times 
the space needed for the compressed sequence itself.
The reported pattern finding rate is below 30 microseconds 
per occurrence, on a DNA collection and patterns of length 10.
This is the first 
self-index based on an algorithm from the LZ77 family.

\section{RLZ-opt in brief}  \label{sec:rlz}

Repetitive data are naturally well handled by compressors from 
the LZ77~\cite{ZL77} family.
The feature common to all those algorithms is to parse the input data into 
a sequence of matches and literals, usually entropy-encoded, e.g. with Huffman 
coding.
An LZ-match (also called a factor) is a reference to an earlier occurrence of 
the same subsequence, 
expressed as the distance (offset) to that earlier subsequence and its length.
If, at the current position, there is no (satisfactory) match, a literal is emitted 
and the compressor moves forward by one symbol.
We note that greedy parsing, i.e., always choosing the longest match, 
is usually a bad strategy.
Parsing the input into matches and literals is a vital factor for the 
compression ratio and the problem of optimal parsing is solved only 
under a simplified assumption of known cost functions 
for encoding match offsets and match lengths~\cite{FNV09}.

RLZ-opt, the algorithm from Kuruppu et al.~\cite{KurPZ11}, follows the LZ77 route, 
but has some features not often met in that family of compressors.
First, it is designed for genomic sequences, where random access to an individual 
sequence (even better, only a small snippet of it) is a welcome feature.
For this reason, one of the sequences in the input collection is chosen 
(and encoded) as the reference sequence, 
while all the other sequences are encoded with relation to the first one, 
but without any cross-references to one another.
Second, matches are found thanks to a suffix array, which is unsual, 
since most LZ77 compressors make use of a hash array (or, more rarely, 
a search tree).
Building a suffix array is relatively slow, but this structure 
facilitates effective non-greedy parsing.

The RLZ-opt algorithm is based on several principles.
A lookahead buffer is used for each considered location in the text, 
which basically means that if the match at position $i+1$ is longer than 
the match at position $i$, a literal may be emitted at position $i$ 
followed by a match.
This idea is however extended, not to a fixed-size buffer, 
but to a buffer whose size changes dynamically, depending on the length 
of the currently longest match found (details can be found in the cited 
work).
Another principle is refraining from LZ-encoding of short matches; 
they are encoded as a run of literals.
This is an idea known in the DNA compression community, 
see e.g.~\cite{MR04}.
Although the match length threshold is fixed (chosen arbitrarily), 
this solution gives a fair boost in compression ratio.
As a last thing, they notice that long and short (i.e., those encoded 
as literals) matches tend to appear alternatively, 
and the offset of the long match can usually be predicted quite well 
from the offset of the previous long match.
Those offsets (match positions) usually form long increasing 
sequences, hence an algorithm for solving the classic 
longest increasing subsequence (LISS) problem is used to detecting 
those matches (called LISS factors), whose offsets are then cheaply encoded.
The LISS factors are often followed by single-symbol factors 
which represent single nucleotide polymorphisms (SNPs) in DNA.
We note that prediction of the next factor position is another 
incarnation of the implicit approximate repeat
detection idea, again known from~\cite{MR04}.

The parsing products in RLZ-opt, e.g. the match lengths and run-of-literals 
lengths, are compacted with Golomb encoding.

\section{Our algorithm}  \label{sec:ouralg}

As said, our algorithm is essentially similar to RLZ-opt~\cite{KurPZ11},
and the main differences are:
\begin{enumerate}
\item in addition to the reference sequence we use extra reference
      phrases from the other sequences for which matches exist,
\item our LZ-parsing is different (details later),
\item our LZ match-finding procedure is based on hashing rather than
      a suffix array, with great benefits for compression speed
      and also adding some flexibility helping to reduce memory
      requirements during the compression,
\item Huffman coding rather than Golomb coding is used in representation
      of various statistics data,
\item compression is performed in blocks (with shared Huffman models),
      to facilitate random access.
\end{enumerate}

Even without the main novelty of our scheme, adding extra reference phrases,
the algorithm (without random access capabilities and segmentation to blocks,
for a fair comparison) produces archives smaller than the Kuruppu et al.\ ones
by from 23\% to 28\%.
We attribute this to the chosen LZ-parsing, which in particular
aggressively looks for a certain class of approximate matches,
and Huffman encoding, not only more compact than Golomb, but also applied
to more byproducts of the compression process.
Incorporating the idea of extra reference phrases increases the reduction to
26\% to 33\%.
First we present the algorithm in the basic form, without the extra 
reference phrases and then explain this enhancement.

\subsection{Basic variant} 
For clarity of exposition we assume that the input data 
are not partitioned into blocks.
One of the input sequences is chosen as the reference sequence 
and all the other sequences are encoded relative to it.
Let us then assume that $T^1$ is the reference sequence, 
and $T^i$ for $2 \leq i \leq r$ are the following sequences that 
will be encoded relatively to $T^1$.
The reference sequence cannot be compressed as effectively as the other ones 
and actually constitutes a substantial part of the final archive.
We first divide the reference sequence into blocks of size 8192 symbols 
and store explicitly start positions of each block in the compressed (output) 
form of this sequence.
It is possible that the start positions for $i$th and $(i-1)$th block 
are equal, which means that the whole block contained only N symbols
(this phenomenon is quite frequent especially on the available human genomes).
Using blocks also enables fast access to data (only local decompression
of the reference sequence is needed).
Then, within each block except for those N-only blocks, we divide 
the reference sequence into triplets over the alphabet of 
size 5 and pack into bytes.
The resulting byte stream is Huffman-encoded,
to prevent inefficiencies on middle-sized runs of N symbols, 
which are not rare either
(we note that a natural means to handle runs of the same symbols, 
the RLE technique, is not so convenient for random access). 
We also hash overlapping subsequences of the reference sequence
(of length $M_1$), to enable further match searches.

In the next phase we process the sequences $T^i$, $2 \leq i \leq r$, 
one by one, scanning them from left to right and looking for 
matches in the reference sequence.
Our parsing strategy is non-greedy but does not mimic the 
lookahead approach known from~\cite{KurPZ11}.
Instead, we adhere to two simple rules: 
a shorter match is chosen if its offset is significantly cheaper 
to encode, 
and it is worth to detect matches with (one or more) single-character 
gaps inside, which are frequent in genomes (a popular form of 
mutation).
Note that the latter idea, although expressed in different terms, 
roughly corresponds to predicting the positions for LISS factors 
in~\cite{KurPZ11}.
Matches with gaps have a gap count limit $k = 2$.
We denote the minimum match length of the first (contiguous) piece 
of a gapped match with $M_1$ and the minimum match extension, i.e., 
the length of each next piece of a gapped match, with $M_2$.

LZ-matches are traditionally represented as a pair: reference offset 
and match length.
We encode offsets as differences between the sequence position 
in the current genome and the matched-to sequence position 
in the reference genome.
This tends to produce relatively small numbers (both negative 
and non-negative) but the extra step, differential encoding of those 
values makes the resulting stream even flatter.

As said, our parsing scheme may prefer a shorter match if its offset 
encoding is cheaper.
The chosen heuristics allows for a match by up to 28 characters shorter 
than the currently most prospective match, if the absolute value of 
its offset in the presented differential form is less than 64 
and this offset property also does not hold for the ``currently best'' 
match.
The constant 64 is related to the byte coding used before further Huffman 
coding; in the former case the offset codeword has 1 byte while in the 
latter case 5 bytes.
We illustrate it with an example.
Let the longest found match be of length 60 and the absolute difference
between the current position (in the current sequence) and the position 
in the referenced sequence (i.e., the beginning of the longest matched 
string) is 2000.
Moreover, the respective absolute difference for the previously encoded 
match is 300.
Apart from that match of length 60 we also have a shorter match, 
of length 40, and the absolute difference between the respective positions 
is only 345 for that shorter match.
Now, since $1700 \geq 64$, $45 < 64$ and $60 - 40 \leq 28$, we prefer 
the shorter match of those two.
A twin heuristic allows to accept a match with a more expensive offset 
if its length is over 28 greater than of the currently best match.

Handling long runs of N symbols deserves special care, 
since it is possible that the reference sequence does not contain them.
To this end, we encode runs of N symbols of length at least $M_1$ 
as a pseudomatch (with the run's actual length and an artificial unique 
offset).

There are four conceptual streams in our solution: match offsets, 
match lengths, literals (those symbols which do not belong to any match) 
and match~/~literal flags.
The latter items are not binary since they also tell the number of gaps 
in a match.
As we limit the number of gaps to 2, the flags are quaternary.
In this way a gapped match is represented by a single offset 
(but the number of encoded match lengths is equal to the number of its 
pieces).
Variable-length byte coding is used for match offsets and match lengths.
The separate byte positions imply separate order-0 Huffman models which 
are responsible for the final compression stage.
For example, the first byte in a match offset has 251 values 
for offset differences from $-125$ to 125, one value telling the offset 
delta is less than $-125$ (followed by 4 extra bytes), 
one value telling the offset delta is greater than 125 
(followed by 4 extra bytes), 
one value denoting an N-run 
and one value for signaling a match to 
a string from the concatenated extra reference phrases
(see later).

Literals are processed like the reference sequence (only without 
dividing them into blocks), 
with packing in triplets into bytes and applying Huffman.
Finally, match~/~literal flags are also byte-packed and submitted to 
yet another Huffman model.

\subsection{Extra reference phrases}
We have noticed that good LZ-compressors are very efficient on our 
data if no restriction to match references is put.
Still, unrestricted set of reference positions to LZ-matches 
prevents random access.
Some compromise has to be found then.
Our solution is to take long enough runs of successive {\em literals}
in $T^i$, $2 \leq i \leq r$, and append them to the reference sequence.
They act as a reservoir for extra matches.
The offset of such a match, as mentioned earlier, has a unique 
1-byte prefix, and what follows is the match position from 
the beginning of the area of extra reference phrases 
(no delta coding used here).
The minimal length of a literal run is $M_3 = 32$.
Note that we detect the literal runs on the fly and attach at the 
end of the extended reference sequence, hence this idea 
does not require an extra pass over data.
In a single pass we cannot be sure, which extra phrases will give a
match for some future sequence, so the value of $M_3$ is chosen
quite arbitrarily, but it cannot be too small;
we assume that the literal run should be longer than the minimal match
length to increase the probability of finding a match to it. 
It is also possible in an extra pass over the compressed data to remove 
the unsuccessfully added extra phrases, but due to the additional 
time we decided not to implement this feature in the current version.

\section{Experimental results} \label{sec:exp}

We have run experiments to evaluate the performance of our algorithm.
The test machine was a 2.4\,GHz Dual-Core AMD Opteron CPU with 16\,GB RAM 
running Red Hat 4.1.2-46, a single core of the CPU was used.
We have implemented our algorithm in C++, and compiled with g++ 4.1.2.

The test collections include the two yeast datasets and 
a dataset of four human genomes, all publicly available 
and used earlier in~\cite{KurPZ11}.

Note there exists a discrepancy in size of the human genomes 
(12066.06\,MB in our tests, 11831.71\,MB in the cited work).
This is due to omitting the N symbols in the reference sequence 
in RLZ-opt.\footnote{S.~Puglisi, private corr., Feb. 2011.}

In our experiments the (very rare, and occurring only in two of the four 
human genomes) symbols different to A, C, G, T, or N are converted to N 
(they denote some more specific kinds of uncertainty in labeling 
nucleotides in the sequencing process).
We found out from the authors that the same methodology was used in~\cite{KurPZ11}.

We tested our algorithm in two variants, where the ``advanced'' one 
includes the idea of extra reference phrases.
The other tested algorithms include 
Comrad 0.2~\cite{KurBSCZ09} and RLZ-opt.
We have not tested the older DNA compressors, XM~\cite{CDAM07} and dna2~\cite{MR04}, 
because their available implementations handle only the 4-symbol alphabet.
In spite of our attempts, we were unable to get a fully working version 
of the LZ-End compressor~\cite{KN10} from the authors;
the program we had in our hands sends to the output only semi-compressed 
data (not interesting from the compression point)
and thus also compression time and especially decompression time cannot 
be honestly measured.
From those reasons we gave up benchmarking~it.

\begin{table}[t]
\begin{center}
\renewcommand{\tabcolsep}{0.0em}
\begin{tabular}{@{\extracolsep{0.8em}}lcccccccc}
\hline
dataset & \multicolumn{4}{c}{S. cerevisiae} & 
          \multicolumn{4}{c}{S. paradoxus} \\
          \cline{2-5}\cline{6-9} 
  &size&ent.&comp.&dec.
  &size&ent.&comp.&dec.\\
  & (MB) & (bpb) & (s) & (s) & 
  (MB) & (bpb) & (s) & (s) \\
\hline
original & 485.87 & 2.18\q & --- & --- & 429.27 & 2.12\q & --- & --- \\\hline
RLZ-opt & \qq9.33 & 0.15\q & \qq310\dqq & \qq\qq2.60 & \q13.44 & 0.25\q & \qq302\dqq & \qq\qq2.56 \\
comrad 0.2  & \q16.50 & 0.27\q & \qq707\dqq & \qq\q41\dqq & \q19.69 & 0.37\q & \qq680\dqq & \qq\q42\dqq \\
\hline
ours, basic & \qq7.18 & 0.118 & \qq\q15.13 & \qq\qq2.97 & \qq9.62 & 0.179 & \qq\q26.52 & \qq\qq2.98 \\
ours, adv. & \qq6.94 & 0.114 & \qq\q12.70 & \qq\qq2.88 & \qq9.01 & 0.168 & \qq\q21.07 & \qq\qq2.76 \\
\hline
\end{tabular}
\end{center}
\caption{Compression results for two repetitive yeast collections.}
\end{table}

\begin{table}[t]
\begin{center}
\renewcommand{\tabcolsep}{0.0em}
\begin{tabular}{@{\extracolsep{1.15em}}lcccccc}
\hline
dataset & \multicolumn{6}{c}{H.\ sapiens} \\
          \cline{2-7}
  &total size & ref.\ seq. size & add.\ seq.\ size  &ent.&comp.&dec.\\
  & (MB) & (MB) & (MB) & (bpb) & (s) & (s) \\
\hline
original & 12066.06 & --- & --- & 2.183 & --- & --- \\\hline
RLZ-opt & \qq703.11 & 639.27 & 63.85 & 0.48\q & 19575\dqq & \q117.73 \\
comrad 0.2 &\q1033.88 & --- & --- & 0.69\q & 32766\dqq & 1157\dqq \\
\hline
ours, basic & \qq730.03 & 702.03 & 28.00 & 0.484 & \qq983.75 & \q152.30 \\
ours, adv. & \qq729.55 & 702.03 & 27.52 & 0.484 & \qq987.34 & \q153.60 \\
\hline
\end{tabular}
\end{center}
\caption{Compression results for four humane genomes.}
\end{table}

On the yeast collections our dominance over the main competitor, 
RLZ-opt, is very significant: 
more than an order of magnitude faster compression
and the archives 
are smaller by 26\% or 33\%, respectively (using our stronger variant).
In decompression, however, RLZ-opt is by about 10\% faster.
Our decompression timings on a machine with a slower CPU stand in contrast 
to those from~\cite{KurPZ11} (which were about three times longer) 
and we can guess this is because our test platform is equipped with a 
15~rpm hard disk.
In fact, it is often reported that for fast LZ77 compressors 
the I/O is a key factor in speed.

On humane genomes the overall picture is seemingly different,
but the extra columns in Table~2 help to explain it: 
our reference sequence is significantly larger after compression, 
since our scheme is simple and random access friendly, 
which is not so in the RLZ-opt where a general-purpose compressor, 
7zip (switches: -mx9 -md512m), with no random access capabilities, 
was used for that.
On the remaining sequences, i.e., in the task of relative compression, 
we are clearly better.
Again, in compression speed our solution dominates (about 20 times faster) 
but RLZ-opt is faster in decompression.

Humane genome sequences are not only large but also much more 
similar to each other than the yeast datasets.
An adverse side-effect of this phenomenon is that there are lots 
of collisions in hashing which make the compression slow.
To mitigate this effect we took a couple of precautions.
On the yeast datasets the minimum match parameters (described 
in the previous section) are set as follows:
$M_1 = 13$, $M_2 = 4$,
while on the humane genomes they are set to:
$M_1 = 20$, $M_2 = 4$.
This, for example, means that the hashed sequences are longer 
on the human genomes and collisions are not that frequent.
Also, on the human sequences we perform the matching on the 
chromosome level rather than on whole sequences
(this concerns all the tested algorithms).
We believe this is a sound approach from the biological point 
and also clearly beneficial for the compression speed and 
memory use requirements.

The parameters $M_1, M_2, M_3$ were chosen experimentally 
but without severe tuning.
Still, it might be interesting to know how their selection affects 
the efficiency of our algorithms.
Recall that $M_3$ corresponds to the minimum length of a 
literal run treated as an extra reference phrase, i.e., setting it 
to a very high value practically turns our advanced variant 
into the basic variant (both in compression ratio and speed), 
which, we believe, is still very competitive.
The parameter $M_2$ is responsible for matches 
with (one-symbol) gaps.
Setting $M_2$ to a much too high value (instead of 4 in all our 
experiments) results in compression loss across all datasets, 
from a few to about 10\%.
The compression speed may also drop moderately.
Finally, $M_1$ cannot be too small, since encoding short matches 
and the triple (offset, length, flag) is costlier than encoding them 
as individual DNA letters.
It seems that the optimum is around $M_1 = 13$ for all datasets.
For human genomes, however, which are much more repetitive than 
other collections of sequences, this threshold had to be raised, 
otherwise the compression speed dropped several times.
According to our preliminary experiments, also for human sequences 
using $M_1 = 13$ gives (slightly) better compression but the price 
is too high.
We admit that manual setting of different values to $M_1$ in our tests 
is a weakness of our algorithm, which can possibly be eliminated by 
some rudimentary quick check on a collection to compress. 

For the yeast datasets we chose as the references
the sequences marked appropriately in their home repository.
It is an interesting problem to find a fast and reliable 
heuristic to select the ``best'' reference sequence 
in the collection.
Our preliminary attempts were unsuccessful but we are going 
to take a closer look at this issue in the future.

For the yeast datasets we chose as the references
the sequences marked appropriately in their home repository.
It is an interesting problem to find a fast and reliable
heuristic to select the ``best'' reference sequence
in the collection.
We found out that choosing the sequence that maximizes
the number of (possibly repeating) subsequences of length $M_1 = 13$
in it works correctly most of the time,
in our experiments failing only in case of one human chromosome.
This heuristics is rather fast, working at about 300\,MB/s in memory,
hence may be a practical choice (however, as said, in our experiments
for Tables 1 and 2 the reference sequences were specified manually).

\section{Conclusions}

The volume of available genomic data grows at an accelerating rate 
and efficient means to store, access and analyze them need to be developed, 
which is addressed in the recent surge of interest in related problems.
We have presented a new efficient data compression scheme for storing DNA 
sequences (genomes) from many individuals of the same species, hence 
sharing lots of similarities.
Experimental comparison with the predecessor of our solution, 
the algorithm of Kuruppu et al.~\cite{KurPZ11}, shows that we are more than 
an order of magniture faster in compression 
and the compression ratio is improved by about 30\%.
The key new ideas are augmenting the reference sequence with extra 
reference phrases taken from the other sequences (without compromising 
decompression speed much), specific LZ-parsing which implicitly 
detects some class of approximate repeats, 
and more compact encoding of the various byproducts of the scheme.
Although not tested yet, we believe our algorithm has a promise 
also for archiving repetitive collections in software version control 
systems.

A number of issues requires further study.
We need to add random access support, which is rather straightforward.
Hash-based LZ-matching, although fast on relatively small data, 
slows down on large inputs, in particular, human genomes.
Some steps to mitigate this problem have been taken in the current 
work but more effort should be put here.
We are going to experiment with hash functions in our setting to 
minimize the number of collisions.
Another aim is to improve the idea of extra reference phrases via 
eliminating those phrases which do not bring overall gain.

An alternative to the idea of extra reference phrases could be using 
two (or more) reference sequences.
In this variant, it is crucial to encode the latter sequence(s) with 
reference to the first one, otherwise on the available datasets, 
with relatively few sequences, a compression loss would be inevitable.
This, of course, poses a burning question about random access to data 
encoded in such a way.
We believe fast random access is possible using the classic rank/select 
operations.
To give a flavor of this idea, let us assume we have two reference sequences 
and the second, $T^2$, of length $n$, is LZ-encoded relative to the first one, 
and we also assume gapless matches for clarity.
We create a (conceptual) bit-vector $B[1, n]$, 
with 1s exactly in the positions where LZ-matches start 
and just after they end.
Now, $rank(B, k)$ is odd iff position $k$ in $T^2$ is inside some match 
and then $k - select(B, rank(B, k))$ tells where exactly within this match $k$ is.
We also need to perform $select$ on the match offsets and lengths 
and have some extra structures to handle prefix sums.
To reduce the overhead of $B$, in reality we use its compressed representation 
(see, e.g.,~\cite{CN08}).

Finally, we note that our algorithm could be used in the inverted index 
from~\cite{CFMPN10} for compressing the text part, 
as a more succinct, but probably slower, alternative to the Re-Pair component.

\section*{Acknowledgments}
We thank Shanika Kuruppu and Simon Puglisi 
for providing us with their software and tips on how to use it.

\end{document}